\documentclass[a4paper,conference]{IEEEtran}
\IEEEoverridecommandlockouts

\IEEEsettopmargin{t}{30mm}
\IEEEquantizetextheight{c}

\usepackage[cmex10]{amsmath}
\usepackage{amssymb,amsthm,mathtools,bm}
\usepackage{enumitem}
\usepackage{algorithm}
\usepackage{algorithmic}
\usepackage{cite}
\usepackage{microtype}
\usepackage[hidelinks]{hyperref}
\usepackage{graphicx}
\usepackage{booktabs}
\usepackage{orcidlink}

\theoremstyle{definition}

\title{Low-Latency Coded Tensor--Matrix Multiplication for Distributed Signal Processing Systems}

\author{
\IEEEauthorblockN{Ahmad Tanha~\orcidlink{0000-0003-3227-8667}}
\IEEEauthorblockA{
EURECOM \& Sorbonne University, 
France\\
Email: ahmad.tanha@eurecom.fr}
\and 
\IEEEauthorblockN{Ali Khalesi~\orcidlink{0000-0002-5815-3611}}
\IEEEauthorblockA{
IPSA \& LINCS Lab,
Paris, France\\
Email: ali.khalesi@ipsa.fr}
}

\begin{document}
\maketitle

\begin{abstract}
Large-scale signal processing systems, including aeronautical and aerospace platforms, increasingly rely on tensor operations, whose distributed execution is constrained by communication, memory, and latency bottlenecks. In this paper, we propose a tensor-aware coded computation framework for distributed mode-1 tensor--matrix multiplication that operates directly on tensor subtensors, avoiding explicit unfolding and preserving multi-dimensional structure.
For a fixed partitioning configuration, the proposed tensor-PolyDot scheme achieves the same recovery threshold, communication cost, worker-side computation, and memory requirements as conventional matrix-based PolyDot schemes. However, by leveraging multivariate encoding, it replaces high-degree univariate interpolation at the fusion node with structured low-dimensional decoding.
This leads to a substantial reduction in decoding complexity and latency, without affecting any other system metric. Additionally, the tensor-based approach improves memory locality and enables parallel decoding across tensor modes, making it well-suited to modern hardware architectures.
Numerical results show decoding speedups proportional to the tensor partitioning factor, reaching up to an order-of-magnitude improvement in practical settings. These advantages make the proposed framework particularly relevant to latency-critical aeronautical and aerospace distributed signal processing systems.
\end{abstract}


\section{Introduction}

Modern signal processing systems increasingly rely on multi-dimensional data representations, where signals are naturally modeled as tensors rather than vectors or matrices. Such representations arise in a wide range of applications, including multi-antenna wireless communications, hyperspectral imaging, video processing, and distributed learning systems~\cite{mozaffari2019tutorial,sidiropoulos2017tensor}. In these settings, tensor operations, such as tensor--matrix products, tensor contractions, and tensor decompositions, form the computational backbone of many signal processing pipelines~\cite{kolda2009tensor}.
Beyond these domains, tensor-based processing is also fundamental in modern aeronautics and aerospace systems, where large-scale, multi-dimensional data must be processed under stringent latency and reliability constraints. For example, airborne radar, synthetic aperture radar, and satellite-based remote sensing inherently generate high-dimensional data capturing spatial, temporal, and spectral structures~\cite{cichocki2015tensor}. Similarly, autonomous aerial vehicles and advanced avionics rely on multi-sensor fusion, integrating heterogeneous data sources, such as radar, LiDAR, and communication signals, into tensor-structured representations that require efficient distributed processing schemes~\cite{khaleghi2013multisensor}.

In such environments, computation is often distributed across onboard processors, edge devices, and ground stations, where communication resources are limited and real-time decision-making is critical. Consequently, decoding at the fusion node frequently becomes a performance bottleneck. This motivates the need for distributed computation strategies that reduce decoding latency without compromising robustness. The proposed tensor-aware coded computation framework addresses this challenge by exploiting multi-dimensional structure to significantly reduce decoding complexity while preserving resilience to stragglers, making it particularly well-suited for next-generation aeronautical and aerospace distributed signal processing systems.

Among tensor operations, mode-$n$ tensor--matrix multiplication plays a central role, enabling dimensionality reduction, feature extraction, and mode-wise filtering. In particular, the mode-1 tensor--matrix product is critical in many applications. For instance, in multi-antenna wireless communication systems, channel state information can be represented as tensors capturing spatial, temporal, and frequency dimensions; processing along the antenna dimension corresponds to mode-1 multiplication and is essential for tasks such as beamforming and channel estimation. Likewise, in hyperspectral imaging, spectral filtering and dimensionality reduction are naturally expressed as tensor--matrix multiplications along specific modes~\cite{cichocki2015tensor}.

As the scale of such data continues to grow, executing tensor operations on a single machine becomes increasingly impractical, necessitating distributed implementations across multiple worker nodes. However, distributed computation introduces several system-level challenges, including communication bottlenecks, limited memory at worker nodes, and the presence of \emph{stragglers}, i.e., slow or unresponsive workers that delay the overall computation~\cite{lee2018speeding}. These challenges are particularly critical in real-time signal processing systems, where latency constraints are stringent.


To address stragglers, \emph{coded computation} techniques have recently emerged as a powerful approach~\cite{li2020codedcomputing,ramamoorthy2020straggler,soleymani2021list}. By introducing structured redundancy, coded schemes enable recovery of the final result from a subset of worker outputs. Notable examples include Polynomial codes, MatDot codes, and PolyDot codes, which achieve efficient tradeoffs between computation, communication, and recovery threshold for distributed matrix multiplication~\cite{yu2017polynomial,dutta2016short,dutta2018unified,posterstruc24}. 
Recent works have also explored structured distributed computation more broadly~\cite{tanha2026thesis}, including sparse tensor factorization~\cite{khalesi2026non} and secrecy-constrained coded matrix multiplication~\cite{tanhaCDMM26}.

Despite these advances, most of the existing coded computation frameworks are inherently matrix-centric: they rely on reshaping or unfolding tensor data into matrices prior to processing. This abstraction overlooks the inherent multi-dimensional structure of tensors, often leading to increased memory traffic, loss of data locality, and additional preprocessing overhead. More importantly, it results in decoding procedures based on high-degree univariate polynomial interpolations, a significant computational bottleneck at the fusion node, given the limited computational resources. This motivates the need for tensor-aware coded computation schemes that directly exploit the multi-dimensional structure to improve system efficiency.

In this work, we address these limitations by proposing a tensor-aware coded computation framework for distributed mode-1 tensor--matrix multiplication that operates directly on tensor subtensors, without requiring explicit unfolding. The key idea is to exploit the multi-dimensional structure of tensors to design \emph{multivariate} encoding schemes, 
which enable decoding via a collection of lower-dimensional
interpolation problems.
This work demonstrates that exploiting tensor structure can substantially alleviate decoding bottlenecks in distributed coded computation.
Compared to conventional matrix-based approaches, the proposed tensor-based framework preserves all fundamental system metrics—such as recovery threshold, communication cost, and worker-side computation and memory—while significantly reducing decoding complexity at the fusion node. 
This reduction directly translates into lower latency, which is critical
in practical distributed and real-time signal processing systems.

\textbf{Our contributions.}
The main contributions of this work are summarized as follows.
\begin{itemize}
\item \textbf{Tensor-aware coded framework:} We propose a coded computation framework for distributed mode-1 tensor--matrix multiplication that operates directly on tensor subtensors, avoiding matrix unfolding and preserving the inherent multi-dimensional structure of the data.

\item \textbf{Multivariate coding for efficient decoding:} We introduce a multivariate encoding and decoding strategy that replaces high-degree univariate interpolation with structured low-dimensional interpolation, reducing decoding complexity by a factor proportional to the tensor partitioning factor.

\item \textbf{Preservation of system guarantees:} We show that the proposed scheme preserves all fundamental system metrics—including recovery threshold, communication cost, and worker-side computation and memory—matching those of matrix-based PolyDot schemes.

\item \textbf{Unified performance characterization:} We provide a comprehensive analysis of recovery threshold, computation, communication, memory, and decoding complexity, revealing a new tradeoff enabled by tensor-aware coding.

\item \textbf{Practical gains in latency:} Through numerical evaluations, we demonstrate that the proposed approach achieves substantial decoding speedups (up to an order of magnitude), translating into significant reductions in end-to-end latency in distributed signal processing systems.
\end{itemize}


\textbf{Notations.}
Scalars are denoted by lowercase or uppercase letters, matrices by bold uppercase letters, and tensors by calligraphic uppercase letters. For $m\in\mathbb{N}$, let $[m]\triangleq\{1,\ldots,m\}$. The mode-1 matrix unfolding of $\mathcal{A}$ is denoted by $\mathcal{A}_{(1)}$, and the mode-1 tensor--matrix product is written as $\mathcal{Y}=\mathcal{A}\times_1\mathbf{B}$. We use $M\triangleq\prod_{n=2}^N I_n$, where $s$ is the partition factor along mode 1, $u_n$ is the partition factor along mode $n$, and $t\triangleq\prod_{n=2}^N u_n$. The recovery threshold is denoted by $K$.

The rest of this paper is organized as follows. In Section~\ref{sec:problem}, we introduce the problem formulation and the distributed computation model. Section~\ref{sec:schemes} presents the coded computation framework and provides a unified analysis of performance metrics, including recovery threshold, computation, communication, and decoding complexity. Section~\ref{sec:numerical} presents numerical results illustrating the advantages of the proposed approach. Finally, Section~\ref{sec:conclusion} concludes the paper and outlines future research directions.

\section{Problem Setup and System Model}\label{sec:problem}

Let $\mathcal{A}\in\mathbb{R}^{I_1\times I_2\times \cdots \times I_N}$ denote an $N$-mode tensor, and let $\mathbf{B}\in\mathbb{R}^{P\times I_1}$ be a matrix. 
We consider the computation of the mode-1 tensor--matrix product
\begin{align}
    \mathcal{Y} = \mathcal{A} \times_1 \mathbf{B},
\end{align}
which produces a tensor $\mathcal{Y} \in \mathbb{R}^{P \times I_2 \times \cdots \times I_N}$.

\paragraph{Element-wise interpretation.}
The mode-1 product corresponds to multiplying the tensor along its first mode. More explicitly, each entry of $\mathcal{Y}$ is given by
\begin{align}
    \mathcal{Y}(p,i_2,\dots,i_N) = \sum_{i_1=1}^{I_1} \mathbf{B}(p,i_1)\,\mathcal{A}(i_1,i_2,\dots,i_N)
\end{align}
for all $p\in[P]$ and $(i_2,\dots,i_N)\in[I_2]\times\cdots\times[I_N]$.

\paragraph{Unfolding interpretation.}
Let $\mathcal{A}_{(1)} \in \mathbb{R}^{I_1 \times M}$ denote the mode-1 matrix unfolding of $\mathcal{A}$, where
\begin{align}
    M \triangleq \prod_{n=2}^N I_n.
\end{align}

Then, the tensor--matrix product can be equivalently written as a matrix multiplication of the form
\begin{align}\label{eq:unfold}
    \mathcal{Y}_{(1)} = \mathbf{B} \mathcal{A}_{(1)} \in \mathbb{R}^{P \times M}.
\end{align}

While this equivalence is useful for analysis, explicitly forming $\mathcal{A}_{(1)}$ may incur significant memory overhead and disrupt data locality. This motivates our tensor-aware approach that avoids explicit unfolding.

\paragraph{Distributed computation setting.}
We consider a distributed system where the input tensor $\mathcal{A}$ and matrix $\mathbf{B}$ are stored at a master node, and the computation is offloaded to multiple worker nodes.

Our objective is to design a coded distributed computation scheme that
\begin{itemize}
\item Minimizes the number of worker responses required for recovery (recovery threshold),
\item Balances computation and memory across workers,
\item Reduces communication overhead, and
\item Minimizes decoding complexity at the fusion node.
\end{itemize}


\paragraph{Partitioning along mode-1.}
We divide the first mode of $\mathcal{A}$ into $s$ equal parts, assuming $s \mid I_1$. Accordingly, the matrix $\mathbf{B}$ is partitioned column-wise into $s$ blocks:
\begin{align}
    \mathbf{B} = [\mathbf{B}_0 \; \mathbf{B}_1 \; \cdots \; \mathbf{B}_{s-1}], 
\quad \mathbf{B}_i \in \mathbb{R}^{P \times \frac{I_1}{s}}.
\end{align}

\paragraph{Partitioning along remaining modes}
For each mode $n \in \{2,\dots,N\}$, we partition the dimension $I_n$ into $u_n$ segments, assuming $u_n \mid I_n$. This results in a total of
\begin{align}
    t \triangleq \prod_{n=2}^N u_n.
\end{align}

subtensors, each of size
\begin{align*}
    \frac{I_1}{s} \times \frac{I_2}{u_2} \times \cdots \times \frac{I_N}{u_N}.
\end{align*}

\paragraph{Block structure in unfolded form}
Under the mode-1 matrix unfolding, each subtensor corresponds to a matrix block of $\mathcal{A}_{(1)}$ of size
\[
\frac{I_1}{s} \times \frac{M}{t}.
\]

\paragraph{Worker computation}
Each worker is assigned encoded combinations of these blocks. After decoding the linear combinations, each worker effectively computes
\begin{align}
    \left(P \times \frac{I_1}{s}\right)
\;\times\;
\left(\frac{I_1}{s} \times \frac{M}{t}\right),
\end{align}
producing an output block of size
\[
P \times \frac{M}{t}.
\]

\paragraph{System model and workflow}
\begin{itemize}
\item \textbf{Encoding (Master):} The master encodes $\mathcal{A}$ and $\mathbf{B}$ and broadcasts them.
\item \textbf{Computation (Workers):} Each worker performs a single local matrix multiplication.
\item \textbf{Decoding (User):} The user reconstructs $\mathcal{Y}$ from a subset of results.
\end{itemize}

\paragraph{Straggler model}
We assume the presence of stragglers, and design the system such that
the result can be recovered from $K$ suitably selected worker responses.

\paragraph{Design objective}
We aim to jointly optimize the recovery threshold $K$, computation and memory at workers, communication cost, and decoding complexity at the user.









\section{Coded Schemes and Performance Analysis}\label{sec:schemes}

In this section, we present two coded computation strategies for distributed mode-1 tensor--matrix multiplication: a baseline matrix-based approach and a proposed tensor-aware scheme. We then provide a unified analysis of their performance in terms of recovery threshold, computation, memory, communication, and decoding complexity.

\subsection{Matrix-Based PolyDot (Baseline \cite{dutta2019optimal})}

A natural approach is to reduce the tensor computation to a matrix multiplication by unfolding $\mathcal{A}$ along mode-1, yielding $\mathcal{A}_{(1)} \in \mathbb{R}^{I_1 \times M}$. The problem then reduces to computing \eqref{eq:unfold}.

PolyDot coding is applied by partitioning $\mathbf{B}$ into $s$ column blocks and $\mathcal{A}_{(1)}$ into $s \times t$ blocks. These blocks are encoded into a single-variable polynomial, where each worker evaluates the polynomial at a distinct point, performs a matrix multiplication, and returns the result.

While this approach achieves efficient distributed computation and robustness to stragglers, it relies on:
\begin{itemize}
\item explicit tensor unfolding, which increases memory traffic,
\item high-degree univariate polynomial interpolation at the decoder.
\end{itemize}

\subsection{Tensor-PolyDot (Proposed)}

We instead operate directly on tensor subtensors without unfolding. The tensor $\mathcal{A}$ is partitioned into $s \times t$ subtensors across its modes, while $\mathbf{B}$ is partitioned along its columns.

The key idea is to encode:
\begin{itemize}
\item mode-1 partitions using a variable $x$,
\item partitions along other modes using variables $y_2,\dots,y_N$,
\end{itemize}
resulting in a \emph{multivariate polynomial encoding}.


Each worker evaluates the encoded inputs at a tuple
$(x,y_2,\ldots,y_N)$, performs a local multiplication, and returns
the result. The fusion node then reconstructs the output using
multivariate interpolation, which decomposes into several
low-dimensional interpolation problems that can be performed in parallel.

This approach preserves tensor structure, avoids data reshaping, and significantly reduces decoding complexity.

\subsection{Recovery Threshold}

For both schemes, the recovery threshold is determined by the total number of distinct polynomial evaluations required to recover all coefficients. This yields $K = (2s-1)t$,
which means that both approaches achieve identical robustness to stragglers.

\subsection{Computation and Memory}

Each worker computes a matrix multiplication of size
\begin{align}
\left(P \times \frac{I_1}{s}\right)
\;\times\;
\left(\frac{I_1}{s} \times \frac{M}{t}\right),
\end{align}
leading to
\begin{align}
    \text{flops/worker} \;\sim\; P \frac{I_1}{s} \frac{M}{t}.
\end{align}

The peak memory requirement per worker, including input and output blocks, is equivalent to
\begin{align}
\text{memory} \;\sim\; 
P\frac{I_1}{s}
+ \frac{I_1}{s}\frac{M}{t}
+ P\frac{M}{t}.
\end{align}

These quantities are identical for both matrix- and tensor-based schemes, since they depend only on partition sizes.

\subsection{Communication Cost}

Each worker returns an output block of size $P \times \frac{M}{t}$. Consequently, the total communication required for the user to recover the result is obtained as
\begin{align}
    \text{Worker} \rightarrow \text{User}:\quad (2s-1)PM.
\end{align}

This quantity is independent of $t$, highlighting that partitioning across additional modes does not increase total communication.

\subsection{Decoding Complexity and Latency}

A key difference between the two schemes lies in the decoding stage, which is detailed below.

\paragraph{Matrix-based decoding (univariate)}
After variable folding, decoding requires interpolation of a univariate polynomial of degree $(2s-1)t-1$, leading to a complexity
\begin{align}\label{eq:cmat}
    C_{\rm mat}\;\sim\; PM (2s-1)^2 t.
\end{align}

\paragraph{Tensor-based decoding (multivariate)}
In the proposed scheme, multivariate decoding decomposes into
low-dimensional interpolations along the coding variables, with total
complexity
\begin{align}\label{eq:cten}
    C_{\rm ten}\;\sim\; PM \Big((2s-1)^2 + \sum_{n=2}^N u_n^2 \Big).
\end{align}

\paragraph{Decoding speedup}
The resulting speedup is
\begin{align}\label{eq:speedup}
    S_{\rm decode}=
\frac{(2s-1)^2t}{(2s-1)^2+\sum_{n=2}^N u_n^2}
    \;\approx\; t
\end{align}
for balanced partitions.

\paragraph{Latency interpretation}
In distributed systems, decoding at the fusion node often lies on the critical path. Thus, reducing decoding complexity directly reduces end-to-end latency. The proposed multivariate scheme achieves substantial latency reduction without affecting other system metrics.

\subsection{System-Level Insights}

The comparison between the two approaches reveals several important system-level observations:

\begin{itemize}
\item \textbf{Decoding efficiency:} The proposed tensor-based scheme reduces decoding complexity by approximately a factor $t$, significantly improving latency at the fusion node.
\item \textbf{Memory locality:} By operating directly on tensor subtensors, the tensor-based approach avoids the explicit data reshuffling associated with tensor unfolding
and can improve memory locality.
\item \textbf{Parallelism:} Multivariate decoding decomposes into multiple smaller interpolation problems, enabling parallel implementation.
\item \textbf{No penalty in other metrics:} Recovery threshold, computation, memory, and communication remain identical to the matrix-based baseline.
\end{itemize}

Overall, the proposed tensor-aware coding framework introduces a new tradeoff dimension: it enables substantial reductions in decoding complexity while preserving all other performance guarantees.

\subsection{Comparison of Coded Schemes}

Table~\ref{tab:comparison} summarizes the key differences between the matrix-based and tensor-based coded computation schemes.

\begin{table}[t]
\centering
\caption{Comparison of Matrix-PolyDot and Tensor-PolyDot Schemes}
\label{tab:comparison}
\renewcommand{\arraystretch}{1.2}

\resizebox{\columnwidth}{!}{%
\begin{tabular}{lcc}
\toprule
\textbf{Metric} & \textbf{Matrix-PolyDot} & \textbf{Tensor-PolyDot (Proposed)} \\
\midrule
Data representation 
& Unfolded matrix $\mathcal{A}_{(1)}$ 
& Native tensor $\mathcal{A}$ \\
Encoding type 
& Univariate polynomial 
& Multivariate polynomial \\
Recovery threshold 
& $(2s-1)t$ 
& $(2s-1)t$ \\
Worker computation 
& $P \frac{I_1}{s} \frac{M}{t}$ 
& Same \\
Worker memory 
& $P\frac{I_1}{s} + \frac{I_1}{s}\frac{M}{t} + P\frac{M}{t}$ 
& Same \\
Communication (total) 
& $(2s-1)PM$ 
& $(2s-1)PM$ \\
Decoding type 
& Univariate interpolation 
& Multivariate interpolation \\
Decoding complexity 
& $PM (2s-1)^2 t$ 
& $PM \big((2s-1)^2 + \sum u_n^2\big)$ \\
Decoding latency 
& High 
& Reduced ($\approx t\times$) \\
Memory locality 
& Poor 
& Good \\
Parallel decoding 
& Limited 
& High \\
\bottomrule
\end{tabular}%
}
\end{table}



















\section{Numerical Evaluation}\label{sec:numerical}

We evaluate the proposed tensor-PolyDot scheme and compare it with the matrix-PolyDot baseline in terms of decoding complexity, runtime, and system-level tradeoffs. Unless otherwise stated, we consider a third-order tensor with
\[
I_1=I_2=I_3=512,\qquad P=256,
\]
and fix the mode-1 partition parameter to $s=8$. The non-shared modes are partitioned using balanced configurations
\[
u_2=u_3=u,\qquad t=u_2u_3=u^2.
\]

We limit the partition factor to $t \le 128$, which corresponds to practical system configurations with moderate numbers of workers and memory constraints.

\subsection{Baseline Configuration}

We first consider a representative setting with $u_2=u_3=4$, yielding $t=16$ and the recovery threshold
\[
K=(2s-1)t=15\cdot 16=240.
\]

The decoding complexities from \eqref{eq:cmat} and \eqref{eq:cten} are
\[
C_{\rm mat}=PM(2s-1)^2t
=2.416\times 10^{11},
\]
and
\[
C_{\rm ten}=PM\big((2s-1)^2+u_2^2+u_3^2\big)
=1.723\times 10^{10}.
\]

This results in a decoding speedup, from \eqref{eq:speedup}, of
\[
S_{\rm decode}=\frac{C_{\rm mat}}{C_{\rm ten}}
\approx 14.0\times.
\]

This example highlights that the proposed tensor-based scheme significantly reduces decoding complexity while preserving identical recovery threshold, computation, memory, and communication cost.

\subsection{Scaling with Tensor Partitioning}

We now examine the impact of the partition factor $t$.

\begin{figure}[t]
\centering
\includegraphics[width=\columnwidth]{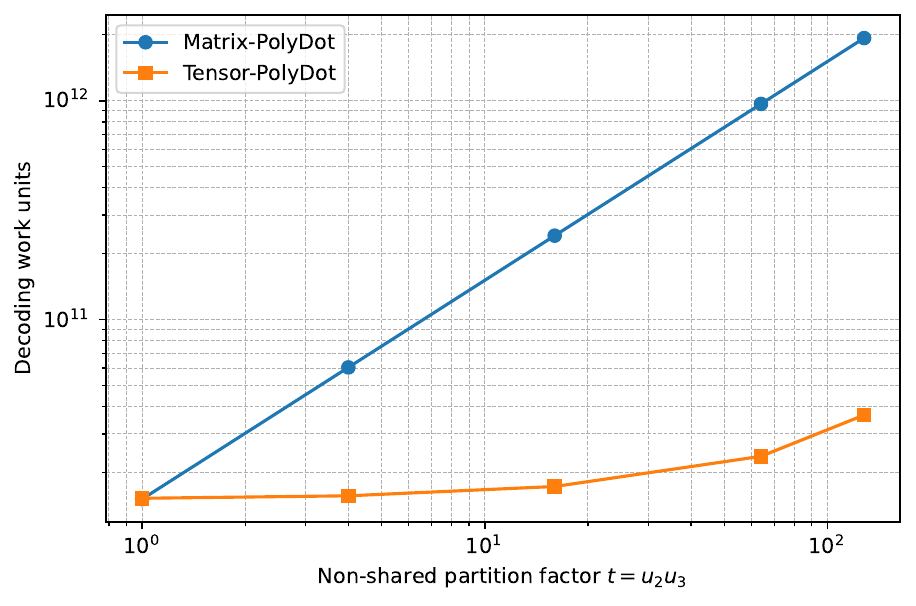}
\caption{Decoding complexity versus non-shared partition factor $t$. Tensor-PolyDot significantly reduces fusion-node decoding complexity compared with matrix-PolyDot.}
\label{fig:decoding_complexity}
\end{figure}

\paragraph{Decoding complexity}
Fig.~\ref{fig:decoding_complexity} shows the decoding complexity as a function of $t$. The matrix-PolyDot scheme scales linearly with $t$, since it requires interpolation of a polynomial of increasing degree. In contrast, tensor-PolyDot grows much more slowly, as decoding is decomposed into multiple lower-dimensional interpolation problems.

\begin{figure}[t]
\centering
\includegraphics[width=\columnwidth]{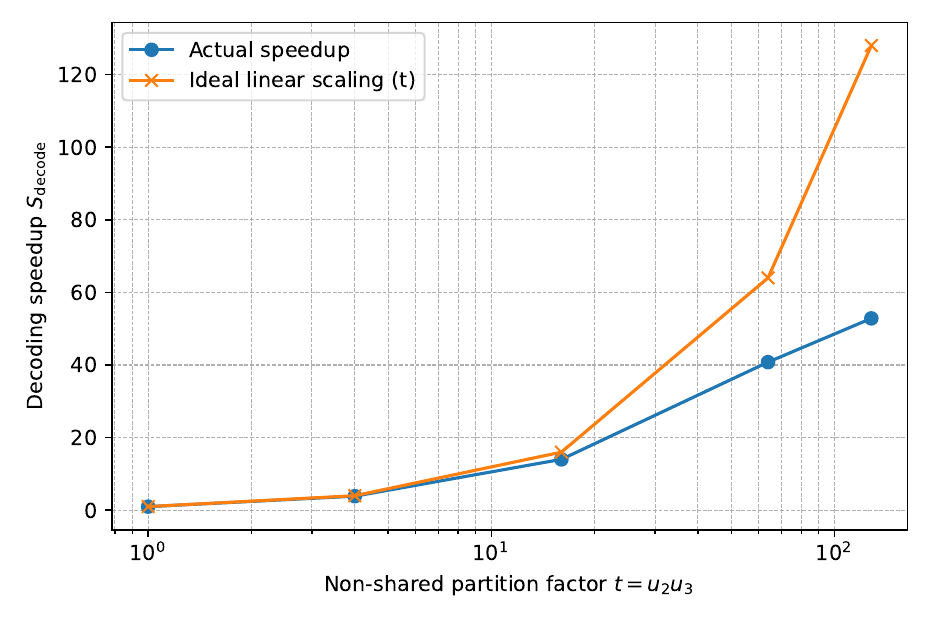}
\caption{Decoding speedup of tensor-PolyDot over matrix-PolyDot.
The achieved speedup approaches the ideal linear scaling with $t$
for balanced partitions.}
\label{fig:decode_speedup}
\end{figure}

\paragraph{Decoding speedup}
Fig.~\ref{fig:decode_speedup} illustrates the decoding speedup in \eqref{eq:speedup}, which increases nearly linearly with $t$ for balanced partitions. For $t=16$, the speedup reaches approximately $14\times$, close to the ideal value $t=16$.

\subsection{Runtime Evaluation and Latency}

To assess practical implications, we translate decoding complexity into runtime using a simple computational model. Specifically, we map the number of decoding operations to execution time assuming a fixed effective interpolation throughput of $5\times 10^9$ operations per second, which captures the memory-bound nature of polynomial interpolation.

\begin{figure}[t]
\centering
\includegraphics[width=\columnwidth]{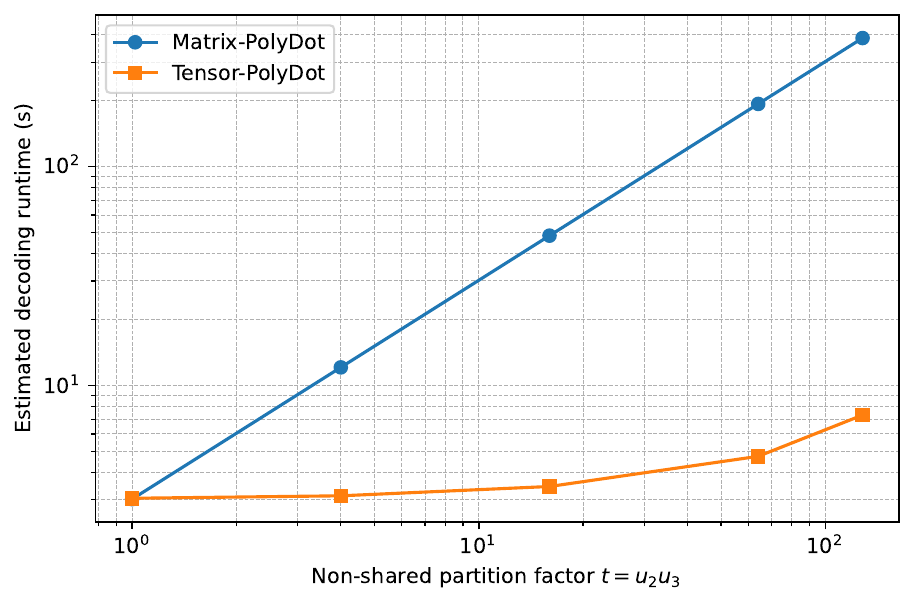}
\caption{Estimated decoding runtime versus non-shared partition factor $t$. Runtime is obtained from decoding complexity assuming an effective CPU throughput of $5\times 10^9$ operations/s.}
\label{fig:runtime}
\end{figure}

\paragraph{Estimated runtime}
Fig.~\ref{fig:runtime} shows the estimated decoding runtime as a function of $t$. The matrix-based scheme exhibits rapidly increasing latency as $t$ grows, whereas tensor-PolyDot maintains significantly lower runtime due to its multivariate decoding structure.

\paragraph{Latency implications}
The results confirm that decoding constitutes a dominant component of end-to-end latency in coded distributed computation systems. By reducing decoding complexity, the proposed tensor-aware scheme effectively alleviates this bottleneck. For moderate partition sizes (e.g., $t=16$), the latency reduction approaches an order of magnitude.

\begin{figure}[t]
\centering
\includegraphics[width=\columnwidth]{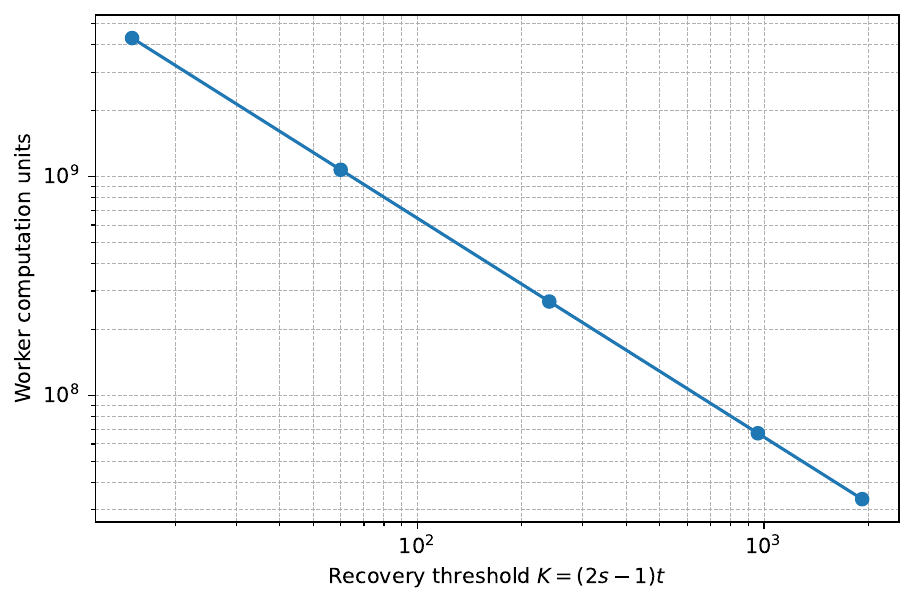}
\caption{Tradeoff between recovery threshold and per-worker computation. Larger $t$ reduces worker computation but increases the number of required worker responses.}
\label{fig:threshold_compute}
\end{figure}

\subsection{Threshold--Computation Tradeoff}

Finally, we examine the tradeoff between recovery threshold and worker computation.

Increasing $t$ reduces per-worker computation, since each worker processes smaller blocks. However, it also increases the recovery threshold
$K=(2s-1)t$,
requiring more worker responses.
Fig.~\ref{fig:threshold_compute} illustrates this tradeoff. As $t$ increases, the system shifts from computation-limited to communication/latency-limited regimes. Therefore, the choice of $t$ should balance available worker resources, memory constraints, and latency requirements.

\subsection{Discussion}

The results reveal three key insights. First, tensor-PolyDot preserves the same recovery threshold and communication cost as matrix-PolyDot for fixed $(s,t)$. Second, the primary gain arises from decoding: multivariate interpolation replaces a single high-degree problem with multiple smaller ones. Third, this reduction directly translates into lower latency, making the proposed approach particularly effective in systems where the fusion node is the performance bottleneck.

Overall, the proposed tensor-aware coding framework improves end-to-end efficiency without imposing additional worker-side or communication overhead, highlighting the benefit of exploiting tensor structure in distributed signal processing systems.

\section{Conclusion}\label{sec:conclusion}

In this work, we proposed a tensor-aware coded computation framework for distributed mode-1 tensor--matrix multiplication. By operating directly on tensor subtensors rather than relying on matrix unfolding, the proposed scheme preserves the inherent multi-dimensional structure of the data and enables a multivariate coding strategy.

We showed that, for a fixed partitioning configuration, the proposed tensor-PolyDot scheme achieves the same recovery threshold, communication cost, and worker-side computation and memory requirements as conventional matrix-based PolyDot approaches. Despite these identical system-level guarantees, the proposed framework significantly reduces decoding complexity at the fusion node by replacing high-degree univariate interpolation with structured multivariate interpolation. This reduction translates directly into lower decoding latency, which is critical in distributed signal processing systems where the fusion node often lies on the critical path.

Our numerical results demonstrate that the decoding complexity can be reduced by a factor proportional to the tensor partitioning factor, yielding substantial speedups, including up to an order-of-magnitude improvement in practical settings. In addition, the tensor-based approach improves memory locality and enables parallel decoding across tensor modes, making it better aligned with modern hardware architectures.

Overall, the proposed framework introduces a new design dimension in coded distributed computing: it enables significant reductions in decoding complexity and latency without incurring any penalty in recovery threshold, communication, or worker-side resources. These properties make it particularly relevant to latency- and resource-constrained aeronautical and aerospace signal processing systems, including airborne sensing, avionics, and satellite-based processing.





\bibliographystyle{IEEEtran}
\bibliography{refs}

\end{document}